%% file: main.tex
\documentclass[12pt,a4paper]{article}
\usepackage{setspace}
\usepackage[vmargin=1.0in,hmargin=0.7in]{geometry}
\usepackage[margin=.4in]{caption}
\input{preamble}

\gdef\@fpheader{\hspace{4mm} Essay written for the Gravity Research Foundation 2021 Awards for Essays on Gravitation}

\title{\bf If time had no beginning}

\author[1]{Bruno Valeixo Bento\footnote{Bruno.Bento@liverpool.ac.uk}}
\author[2]{Stav Zalel\footnote{stav.zalel11@imperial.ac.uk}}
\affil[1]{\small Department of Mathematical Sciences, University of Liverpool, L69 7ZL, U.K.}
\affil[2]{Blackett Laboratory, Imperial College London, SW7 2AZ, U.K.}

\usepackage[hidelinks]{hyperref}

\begin{document}

\noindent{\footnotesize\scshape\@fpheader}\par
{\let\newpage\relax\maketitle}
\maketitle

\abstract{General Relativity traces the evolution of our Universe back to a Big Bang singularity. To probe physics before the singularity---if indeed there is a ``before''---we must turn to quantum gravity. The Causal Set approach to quantum gravity provides us with a causal structure in the absence of the continuum, thus allowing us to go beyond the Big Bang and consider cosmologies in which time has no beginning. But is a time with no beginning in contradiction with a passage of time? In the Causal Set approach, the passage of time is captured by a process of spacetime growth. We describe how to adapt this process for causal sets in which time has no beginning and discuss the consequences for the nature of time.}

\section{Time and Causal Sets}
Did time ever begin? It is hard to decide which answer is more unsettling: the idea of an infinite past with no beginning or the concept of such a beginning---the birth of the Universe. 
Stephen Hawking proved that General Relativity (GR) breaks down at a Big Bang singularity, but left open the possibility that the Big Bang is \textit{not} the beginning of time but rather that it was preceded by a quantum gravity era which cannot be captured by GR \cite{Hawking:2014xx}. The question of the beginning of time must therefore be addressed within a theory of quantum gravity.
%
%

\begin{figure}[h]
    \centering
    \includegraphics{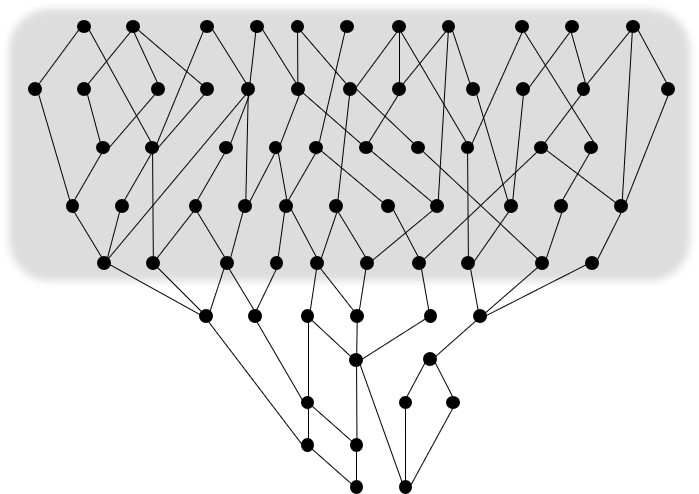}
    \caption{
    A causal set. Elements are represented as nodes and the order is indicated by the edges: element $x$ precedes element $y$ if and only if there is an upward-going path from $x$ to $y$. The portion of the causal set which lies in the shaded region is well approximated by a continuum spacetime (physics in this region is captured by GR). The remainder of the causal set forms the quantum gravity era preceding the Big Bang singularity.}
    \label{fig:continuum}
\end{figure}

Causal Set Theory is an approach to quantum gravity which postulates that spacetime is fundamentally discrete and takes the form of a \textit{causal set}, a partial order whose elements are the indivisible ``atoms'' of spacetime \cite{Bombelli:1987aa,Surya:2019ndm}. The partial order is interpreted as a temporal order, so that the \textit{past} of an element is formed of all the elements which precede it in the partial order. Thus the causal set furnishes a causal structure---a notion of before and after---in the absence of the continuum, allowing us to contemplate whether there was anything ``before'' the Big Bang (Fig.\ref{fig:continuum}) \cite{Dowker:2017zqj}.



Naively, we may consider the continuum spacetime of GR to emerge from an underlying causal set via a large (length) scale approximation \cite{Dowker:2005tz}. But quantum mechanics suggests that reality is better described as a superposition of causal sets. A quantum theory of causal sets will ultimately be formulated as a sum-over-histories---a ``path integral" of sorts---with the causal set playing the role of ``history'' or ``spacetime configuration'' \cite{Sorkin:1997gi,Sorkin:2006wq,Sorkin:1994dt}. Assigning a weight to each history in the sum is the problem of causal set dynamics.


Much of the effort towards obtaining a dynamics for causal sets has been guided by the paradigm of \textit{growth dynamics} which states that the weight/action emerges from a fundamental physical process in which the causal set comes into being \textit{ex nihilo}. This notion of \textit{becoming}, the idea that a causal set grows element by element, further allows the passage of time to be captured by physics: an instantaneous moment---a \textit{now}---corresponds to the birth (not to the existence) of an element \cite{Sorkin:2007hga,Dowker:2014xga,Dowker:2020qqs}.

Kinematically, causal sets can provide a cosmology in which time has no beginning---namely, a causal set in which every element has an infinite past. But are such past-infinite causal sets compatible with the heuristic of growth and becoming? If not, we may be forced to choose between a passage of time and a beginningless time.
 
\section{Growth Dynamics: Sequential vs Covariant}



\begin{figure}[h]
    \centering
    \includegraphics[width=0.8\textwidth]{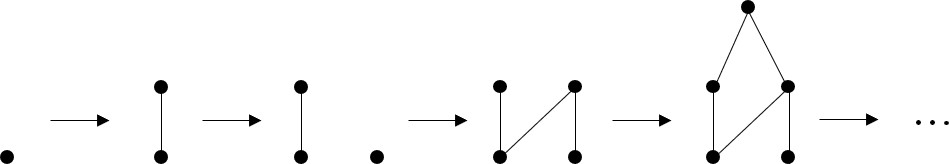}
    \caption{Sequential growth. Elements are born in a total order, one after the other. The total order of births is unphysical (pure gauge).}
    \label{fig:growth}
\end{figure}

In its fully-fledged form, the growth process will be a quantum phenomenon \cite{Sorkin:2011sp,Criscuolo:1998gd,Dowker:2010qh,Surya:2020cfm} but at this stage of development of Causal Set Theory, growth dynamics are classical stochastic processes which generate infinite causal sets. Thus far, the most fruitful growth dynamics are the Classical Sequential Growth (CSG) models \cite{Rideout:1999ub} in which, starting from the empty set, a single element is born at each stage (Fig.\ref{fig:growth}). The ordering of each new-born element with respect to the already-existing elements is determined probabilistically according to each model but always satisfies the constraint that a new-born element cannot precede an already-existing one, ensuring a consistency between the interpretation of the partial order as a temporal order and of the birth of elements as the passage of time.

%

\begin{figure}
     \centering
     \begin{subfigure}[b]{0.25\textwidth}
         \centering
         \includegraphics[width=\textwidth]{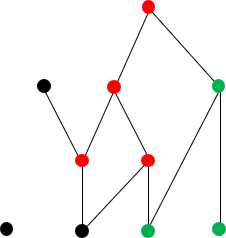}
         \caption{}
         \label{fig:causalsetwithsets}
     \end{subfigure}\hspace{15mm} %
     \begin{subfigure}[b]{0.4\textwidth}
         \centering
         \includegraphics[width=\textwidth]{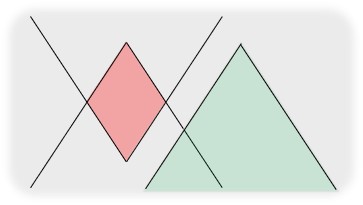}
         \caption{}
         \label{fig:onestem}
     \end{subfigure}
 \caption[]{
    $(a)$ Stems and convex sets. The green ``portion'' is a stem because it is finite and it contains its own past (\textit{i.e.} the past of each of its elements). The red ``portion'' is \textit{not} a stem because it does not contain its entire past (\textit{e.g.} it does not contain the green elements), but it is a convex set because it contains all the elements which lie \textit{in between} its elements in the partial order. The black ``portion'' is neither a stem nor a convex set.  $(b)$ Continuum analogues of stems and convex sets. A stem corresponds to any union of past lightcones whose total spacetime volume is finite. A causal set with no beginning contains no stems, just like a geodesically complete spacetime contains no past lightcone of finite spacetime volume. A convex set is a generalisation of the intersection of a past lightcone with a future lightcone. }
    \label{fig:stem}
\end{figure}

Our individual experience of the passage of time as a linear, totally ordered sequence of events is reflected in the sequential nature of the CSG models where elements are born in a sequence, one after the other. But this familiar notion of becoming is too simplistic to capture the intrinsic partial order/causal structure, since the total order acts as a gauge global time. The struggle between the gauge formulation of sequential growth and the gauge-independent nature of the physical world (cf. local coordinates and general covariance in GR) is resolved by identifying gauge-independent observables. The role of observables is played by \textit{stems}, finite ``portions'' of a causal set which contain their own past (Fig.\ref{fig:stem}). In other words, in CSG models the growing causal set is fully determined by its stems \cite{Brightwell:2002yu,Brightwell:2002vw,Dowker:2005gj}.

The CSG models are toy models of quantum cosmology but their original formulation shies away from the question at hand---whether time began---since the condition which prohibits new-born elements from preceding already-existing ones means that the growth process can only produce causal sets in which time has a beginning.
Loosening this restriction by allowing new-born elements to precede already-existing ones opens a new avenue for causal set cosmology in which the problem of the beginning of time can be formalised \cite{convexcovtree}. But how should this new form of growth, in which the order of births is incompatible with the partial order, be understood? If element $x$ precedes element $y$ in the temporal partial order, what could it possibly mean for $y$ to be born before $x$? It is hard to see how the growth can be considered a real physical process in this modified framework. Is a time with no beginning inherently incompatible with 
the notion of becoming?

The missing piece that may reconcile a beginningless time with a physical growth process is to replace our intuitive notion of sequential becoming with \textit{asynchronous becoming} where elements are born in a partial (not a total) order \cite{Sorkin:2007hga,Dowker:2014xga,Dowker:2020qqs}. What does it mean for elements to be born in a partial order? Through the lens of our largely sequential experience, asynchronous becoming may sound more like a fantastical riddle than a description of physical reality. It is the role of mathematics to make sense of notions which lie beyond our everyday experience, and it may be that new mathematics is what is needed to better understand asynchronous becoming and its consequences for the nature of time. 

\begin{figure}[h]
    \centering
    \includegraphics[width=0.85\textwidth]{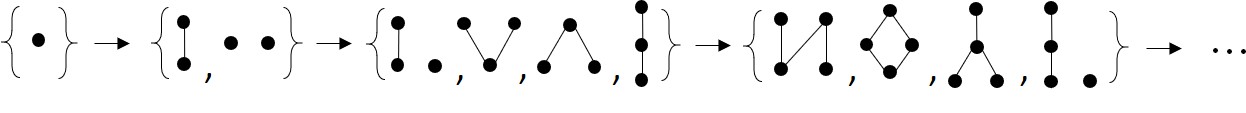}
    \caption{Covariant growth. The growth process does not keep track of the birth of individual elements but rather of the stems in the growing causal set. The $n^{th}$ set in the sequence contains all the causal sets which have cardinality $n$ and are stems in the growing causal set, so that after $n$ steps all the stems of cardinality $\leq n$ are determined.}
    \label{fig:covtree}
\end{figure}

\textit{Covariant growth} is an alternative to sequential growth which may contain the seed of asynchronous becoming \cite{Dowker:2019qiz,Zalel:2020oyf}. In its original formulation, covariant growth only produces causal sets in which time has a beginning.  Taking its cue from the CSG models, covariant growth assumes from the outset that a causal set spacetime is fully described by its stems (\textit{i.e.} that causal sets which share all the same stems are physically equivalent). Thus, in contrast to sequential growth, covariant growth does not keep track of individual element births but only of the stems contained in the growing causal set. The growth process can be illustrated as a sequence of sets, where the $n^{th}$ set in the sequence contains all the causal sets which have cardinality $n$ and are stems in the growing causal set (Fig.\ref{fig:covtree}). When the process runs to completion (in the $n\rightarrow\infty$ limit) all stems are determined, thus fully determining the causal set spacetime grown in the process.

While the process of becoming is explicit in sequential growth, it is implicit or ``vague'' \cite{Wuthrich:2015vva} in covariant growth (\textit{e.g.} at any finite stage of the growth process, one cannot say which portion of the causal set has already come into being). But if there is a process of becoming which can be associated with covariant growth, then it may be that it is this quality of vagueness which embodies asynchronous becoming and thus allows us to reconcile the passage of time with a beginningless time in Causal Set Theory.


\section{Causal sets with no beginning}

Covariant growth can be modified to accommodate growth of causal sets in which time has no beginning. The key is identifying the observables pertaining to these causal sets. A causal set with no beginning contains no stems, since if a portion of the causal set contains its own past then it must contain infinitely many elements, while stems have finite cardinality by definition. Instead, the role of observables is played by \textit{convex sets}, ``portions'' of a causal set which, whenever they contain a pair of elements $x$ and $y$, contain all elements which lie between $x$ and $y$ in the partial order (Fig.\ref{fig:stem}). If finite convex sets encode all that is physical in a causal set, then we can adapt the covariant growth process for past-infinite causal sets simply by replacing stems with convex sets \cite{convexcovtree}.  This new formulation of covariant growth keeps track of convex sets contained in the growing causal set. At stage $n$, all convex sets of cardinality $n$ are fixed so that in the $n\rightarrow \infty$ limit the causal set spacetime is fully determined.

The significance of this new covariant formalism is twofold. First, this process is capable of growing all kinds of causal sets: in some time begins, in others it does not. Thus, whether time has a beginning or not is no longer a choice hardwired into our construction but rather a question which we can ask of the dynamics. Second, the implicit nature of the growth means that there is no immediate contradiction between the process of becoming and the past-infinite nature of a growing causal set. It will be up to future work to decide whether covariant growth can really be interpreted as a physical growth of past-infinite causal sets; whether there is a yet unknown formalism which better encompasses asynchronous becoming and in doing so captures the passage of a beginningless time; or whether the physics of passage dictates that time must have a beginning.

\vspace{2mm}
\noindent \textbf{Acknowledgments:} The authors are indebted to Fay Dowker for guidance and collaboration on the work presented in this essay.

\bibliography{covtree}{}
\bibliographystyle{unsrt}

\newpage

\end{document}

%% file: preamble.tex
\usepackage{authblk}
\usepackage{graphicx}
\usepackage{url}
\usepackage{amstext}
\usepackage{amsmath}
\usepackage{amsfonts}
\usepackage{subcaption}
\usepackage{mathtools}
\usepackage{float}
\usepackage{fancyhdr}
\usepackage{amsthm}
\usepackage{multirow}
\usepackage{cite}
\usepackage{placeins}
\usepackage{makecell}

\usepackage{tikz}

\usepackage{caption}
\usepackage{soul}

\usepackage{verbatim}